\documentclass[default]{svmult}

\usepackage{mathptmx}       
\usepackage{helvet}         
\usepackage{courier}        
\usepackage{type1cm}        
\usepackage{makeidx}         
\usepackage{graphicx}        
\usepackage{multicol}        
\usepackage[bottom]{footmisc}
\usepackage{epstopdf}

\usepackage{color}

\makeindex             

\begin{document}

\title*{A Cooperative Emergency Navigation Framework using Mobile Cloud Computing}
\author{Huibo Bi and Erol Gelenbe}
\institute{Imperial College London
Department of Electrical and Electronic Engineering
Intelligent Systems and Networks Group, \email{{huibo.bi12,e.gelenbe}@imperial.ac.uk}}
%
%
\maketitle

\abstract{The use of wireless sensor networks (WSNs) for emergency navigation systems suffer disadvantages such as limited computing capacity, restricted battery power and high likelihood of malfunction due to the harsh physical environment. By making use of the powerful sensing ability of smart phones, this paper presents a cloud-enabled emergency navigation framework to guide evacuees in a coordinated manner and improve the reliability and resilience in both communication and localization. By using social potential fields (SPF), evacuees form clusters during an evacuation process and are directed to egresses with the aid of a Cognitive Packet Networks (CPN) based algorithm. Rather than just rely on the conventional telecommunications infrastructures, we suggest an Ad hoc Cognitive Packet Network (AHCPN) based protocol to prolong the life time of smart phones, that adaptively searches optimal communication routes between portable devices and the egress node that provides access to a cloud server with respect to the remaining battery power of smart phones and the time latency.}


\section{Introduction}
\label{sec:intro}

With the rapid development of sensor technology, emergency evacuation in built environments requires time-critical response with respect to multi-domain sensing, data interpreting and information transmitting.  WSN based emergency management systems cannot provide optimal solutions due to their limited computing capability, battery power and storage capacity. Although the evolution from homogeneous architecture with functionality-identical sensors to heterogeneous architectures with separated sensing and decision subsystems makes WSNs based systems more energy-efficient and fault-tolerant, most decision supporting subsystems that consist of lightweight decision nodes still suffer from resource restriction problems.

Owing to the high processing power and low risk level, cloud-computing has become a dominating technology and tends to revolutionise the emergency management landscape. Accompanying this tendency, a new interest has been aroused to consider smart phones as simple clients for the back-end Cloud due to their sensing ability and popularity \cite{gelenbe2013future}. For example, Ref. \cite{chu2011real} presents an emergency navigation system based on smart phones and active temperature RFID sensor tags to calculate evacuation routes and  intensive computations are offloaded to a cloud server. Due to the limitation of indoor positioning and unavailability of GPS, in \cite{ahn2011rescueme} a smart phone assisted system to locate evacuees with pedometry-based localization and image-based positioning is suggested. The cloud based server can obtain the position of individuals by matching the image snapshots uploaded by evacuees and then provide uncrowded routes for users. However, current cloud enabled emergency response systems with the aid of portable devices do not consider the impact of the significant energy consumption in the client side during the communication process. On the other hand smart phones using the UMTS protocol can easily come under attack \cite{NEMESYS1,NEMESYS2} including various forms of denial of service attacks \cite{Storms}.

Much research has demonstrated cooperative strategies have a significant positive influence on multi-agent systems and related studies have emerged for decades as organizational paradigms in agent based structures \cite{horling2004survey} and autonomous search by large scale robotic systems \cite{gelenbe1997autonomous,Cao,Cao1}. However, most research on emergency evacuation focuses on developing different models \cite{zheng2009modeling} to simulate the crowd or coordinated behaviours such as kin behaviour \cite{yang2005simulation}. Although previous work  \cite{gorbil2013resilient} has proposed a resilient emergency support system (ESS) with the aid of opportunistic communications \cite{pelusi2006opportunistic} to study the impact of ``passive'' cooperation among evacuees which exchange emergent messages with other civilians within the communication range, no active mechanism is used to improve the information dissemination efficiency.

Thus, in this paper we propose a cloud based emergency navigation system which uses a cooperative strategy to egress evacuees in loose clusters, and we also present a power-aware quality of service (QoS) metric to prolong the life time of smart phones used in an evacuation to connect to the access point(s) of the cloud system that carries out intensive computations.

The remainder of the paper is organized as follows: In Section \ref{CPN} we recall the concept of Cognitive Packet Networks, and then describe related algorithms in Section \ref{System}. The simulation model and assumptions are introduced in Section \ref{Sim} and results are presented in Section \ref{Results}. Finally, we draw conclusions in Section \ref{Conclu}.

\section{Cognitive Packet Network} \label{CPN}

The Cognitive Packet Network (CPN) introduced in \cite{gelenbe2001towards,gelenbe2004self,gelenbe2009steps} was originally proposed for large-scale and fast-changing packet networks. By employing cognitive or smart packets (SPs), CPN can discover optimal routes rapidly and heuristically and realise continuous self-improvement. Contrary to conventional routing protocols, in the CPN, intelligence is realised by using Random Neural Networks \cite{gelenbe1989random,gelenbe1990stability,gelenbe1993learning,RNN2008,gelenbe2001simulations,Natural} and is constructed into smart packets other than protocols. Hence, cognitive packets can discover optimised routes with their predefined goals and improve QoS by learning from their own investigations and experience from other packets.

The AHCPN \cite{gelenbe2004power} is a variant of the original CPN. It can replace most of broadcast transmissions with unicast transmissions and adapt to highly dynamic Ad hoc network environments rapidly.

\section{The Cloud-enabled Emergency Navigation framework} \label{System}

The framework contains a user layer which is composed of evacuees with portable devices and a Cloud layer which performs intensive computations. In the following subsections, we concentrate on the coordinated emergency navigation algorithm and the energy efficient protocol due to the space constraints.

\subsection{Coordinated Emergency Navigation}

Evacuating in a cooperative manner can increase the probability for evacuees to obtain assistance when at risk and reduce energy utilization of smart phones by relaying data. In this section, we leverage the SPF to cluster evacuees into groups during an evacuation. The initial SPF algorithm \cite{reif1999social} consists of global control forces and local control forces. The global control forces coordinate the individuals and determine the distribution of the individuals while the local control forces dominate the personal behaviours. In our treatment, we replace the local control forces with the CPN based algorithm to control the individual behaviour of evacuees and use the global forces to regulate intra-group behaviours. Details of the CPN based emergency evacuation algorithm can be seen in \cite{BiDesmetGelenbeISCIS2013} and \cite{bi2014routing}. The force between two evacuees can be calculated from the following equation:
\begin{equation}
f(r) = -\frac{c_{1}}{r^{\sigma_{1}}} + \frac{c_{2}}{r^{\sigma_{2}}}
\end{equation}
where $c_1$, $c_2$, $\sigma_{1}$ and $\sigma_{2}$ are constants and $r$ is the distance between two evacuees. Term $-\frac{c_{1}}{r^{\sigma_{1}}}$ presents the repulsive force while $\frac{c_{2}}{r^{\sigma_{2}}}$ depicts the attractive force. We assume that each civilian can only be affected by other civilians within 20 meters. If the distance between two civilians is smaller than 7 meters, a repulsive force will be generated, otherwise, a attractive force will be produced. To achieve this, we set $c_{1}$ to 20, $c_{2}$ to 15, $\sigma_{1}$ to 0.9478 and $\sigma_{2}$ to 0.8, respectively. This ensures the civilians to evacuate in loose groups without increasing the level of congestion.

To combine the SPF and the CPN based algorithm, we use a simple scheme to randomly choose either the decision of the SPF or the CPN as the next decision at a time. When a SPF related decision is chosen, we adopt the associated neighbour node which has the most matched direction with the resultant force as the next hop.

\subsection{Power-aware Protocol}

As the front-end component of the cloud enabled framework, the operational time of smart phones becomes a bottleneck of the system due to the large amount of energy consumption during information exchanges. Because the remaining battery power of the portable devices yields normal distribution when an emergency event occurs and the power consumption of different communication modes varies, it is not a optimal strategy for each smart phone to exchange information with the Cloud through 3G directly. To realise energy efficiency and maximise the average battery lifetime of smart phones, we employ an AHCPN based algorithm to relay sensory data before ultimately uploading to the Cloud. The AHCPN based algorithm is deployed on the smart phones to search power-saving paths to convey sensory data to the Cloud. The QoS criterion we used is inspired by the energy aware metric in \cite{gelenbe2004power}. Here we also employ the ``path availability'' notion and construct the metric as follows.

\begin{equation}
G_{ed} = \alpha \prod_{i=0}^{n-1}P_a(\pi(i),\pi(i+1))\left\{\sum_{i=0}^{n-1}D(\pi(i),\pi(i+1))\right\}
\end{equation}
where $\pi$ represents a particular path, $n$ is the number of nodes (smart phones) on the path $\pi$, and $\pi(i)$ is the $i$-th node on the path $\pi$. Term $P_a(\pi(i),\pi(i+1))$ is the availability of the edge between $\pi(i)$ and $\pi(i+1)$. $D(\pi(i),\pi(i+1))$ is the delay cost for a packet to transmit from $\pi(i)$ to $\pi(i+1)$. Term $\alpha$ is a constant that coordinates the relation between the path availability and delay.

Path availability is affected by the remaining battery power of a smart phone and the estimated power consumption of transmitting a piece of information:
\begin{equation}
P_a(\pi(i),\pi(i+1)) = \frac{B_i^{C}}{B_i^{C} - B_i^{U}}
\end{equation}
where $B_i^{C}$ represents the current remaining battery power of a node (smart phone) $\pi(i)$  and $B_i^{U}$ depicts the power utilisation at node $\pi(i)$ to convey a certain piece of information. If the potential power utilisation is larger than the remaining battery, this node will be excluded.

\section{Simulation Model and Assumptions}
\label{Sim}

\begin{figure}[ht]
\centering
\includegraphics[width=0.5\textwidth]{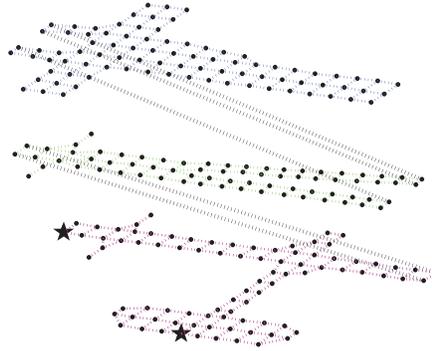}
\caption{Graph representation of the building model: vertices are positions with landmarks that evacuees can easily identify their locations by uploading snapshots and matching them with pre-known landmarks which are stored in cloud severs, edges are physical links that evacuees can move inside the building. The two black stars on the ground floor mark the position of the building's exits.}
\label{fig: graph3D}
\end{figure}

Since mathematical models \cite{Acta79} cannot handle the full complexity that is encountered during an emergency, the performance of the proposed algorithms is evaluated in fire-related scenarios based on the Distributed Building Evacuation Simulator (DBES)\cite{Avgoustinos,Dimakis}. The building model is a three story canary wharf shopping mall as shown in Fig. \ref{fig: graph3D}. Initially, evacuees are randomly scattered in the building and are equipped with smart phones with random remaining battery power. When the battery power of a smart phone is depleted, the evacuee will wander or follow other evacuees in the line of sight to exits.

We assume that no related sensors are pre-installed in the built environment. Hence, hazard information and location of civilians will be collected by mobile phones with built-in cameras and analysed in the Cloud. We also hypothesis that in-door communication infrastructures such as Wi-Fi access points are unavailable but a few cloud access points can be rapidly deployed between the emergency location and the Cloud when a hazard occurs. According to literature \cite{carroll2010analysis,haas2012realistic,balasubramanian2009energy,kalic2012energy}, the energy consumption model of an individual smart phone is shown in Table \ref{table: energymodel}.

\begin{table}
    \begin{center}
        \begin{tabular}{| r | c | c | c |}
            \hline
            Communication model & Download & Upload & Signal Rate \\ \hline
            3G ~~~~~~~~~~~~~ & $0.001224x$ & $0.0003375x$ & 2 Mb/s \\ \hline
            Bluetooth ~~~~~~ & $0.0001377x$ & $0.00012012x$ & 1 Mb/s \\ \hline
            \hline
        \end{tabular}
        \caption{The energy consumption of a smart phone in Joule with regard to the transferred
data $x$ in byte and the related signal rate.}
        \label{table: energymodel}
    \end{center}
\end{table}

\section{Results and Discussion}
\label{Results}

To evaluate the navigation algorithm which combines CPN and SPF (CPN\&SPF) as well as the AHCPN based energy efficient protocol, we design an experiment which involves 4 scenarios with 30, 60, 90, 120 evacuees, respectively. The Dijkstra's shortest path algorithm and CPN based algorithm with time metric (CPNST) \cite{bi2013routing} are performed for comparison purpose.

\begin{figure}[ht]
\centering
\includegraphics[width=0.8\textwidth]{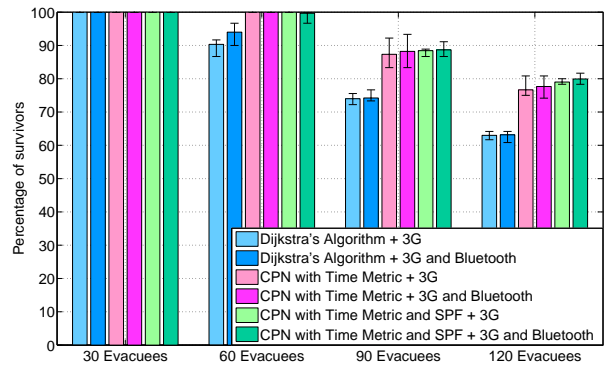}
\caption{Percentage of survivors for each scenario. The results are the average of five randomized simulation runs, and error bars shows the min/max result in any of the five simulation runs.}
\label{fig: numberofSurvivors}
\end{figure}

Fig. \ref{fig: numberofSurvivors} indicates that both CPNST and CPN\&SPF achieve more survivors than Dijkstra's algorithm due to the congestion-ease mechanisms in both CPNST and SPF. Compared with CPNST, CPN\&SPF performs slightly better in densely-populated scenarios because CPN\&SPF organises evacuees as loose clusters and is easier to balance the power utilisation among smart phones. As a result, the probability of being trapped due to the depletion of smart phones is decreased. Furthermore, the coordinated behaviour increases the probability for an evacuee with depleted smart phone to follow other evacuees other than wander in the building.

\begin{figure}[ht]
\centering
\includegraphics[width=0.8\textwidth]{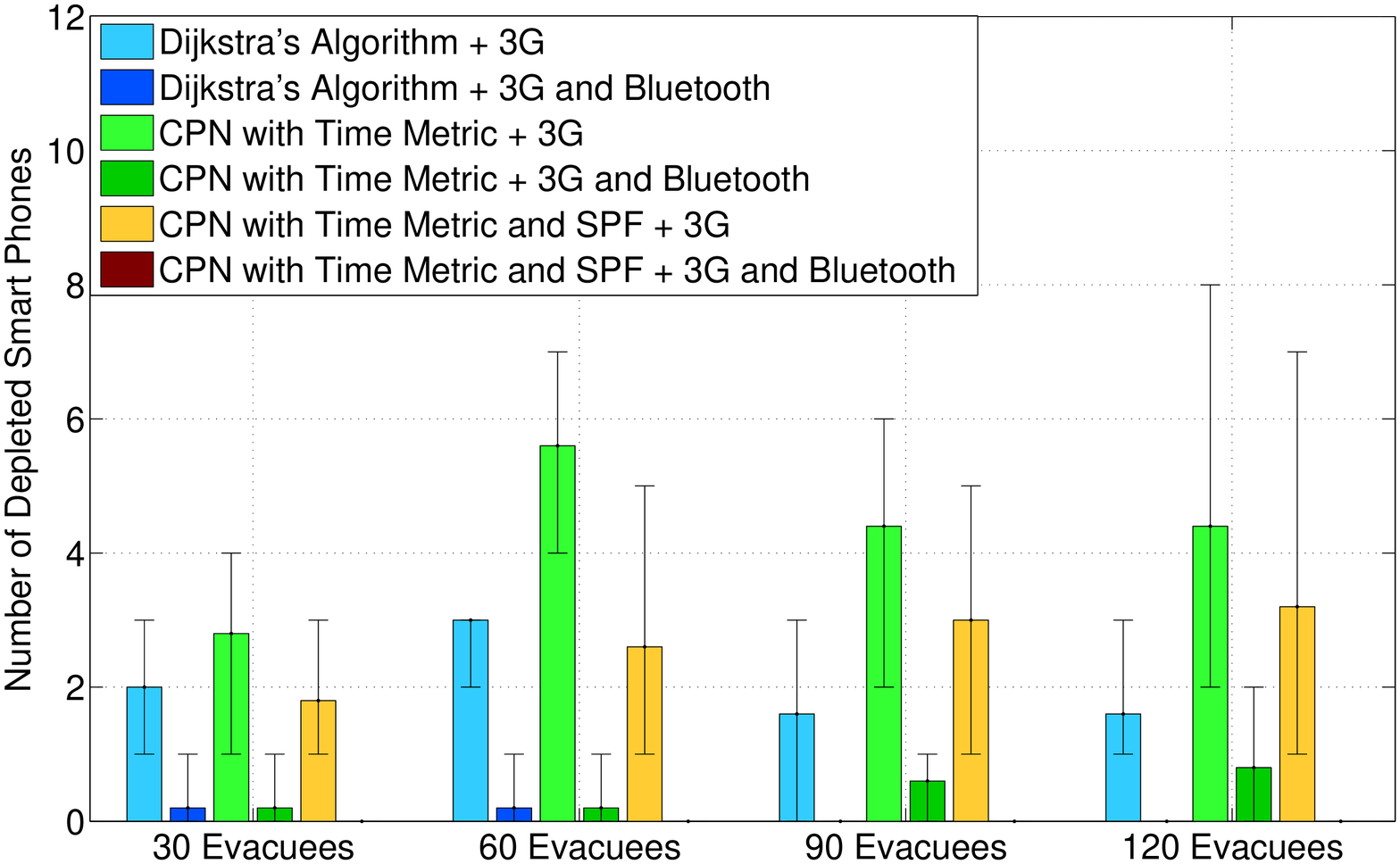}
\caption{The number of drained smart phones of five iterations. The error bars represent the min/max values found in the five simulations.}
\label{fig: deadsmartphones}
\end{figure}

As can be seen from Fig. \ref{fig: deadsmartphones}, Dijkstra's algorithm achieves least number of drained smart phones in 3G mode. This is because both the CPNST and CPN\&SPF  do not follow the shortest path. Hence, evacuees will traverse more landmarks and upload more photos to the Cloud. However, by employing the AHCPN protocol, the number of depleted smart phones decreases significantly for all three algorithms. Moreover, there is no drained smart phone when combining the CPN\&SPF algorithm with the AHCPN protocol. This confirms that routing evacuees in loose groups can contribute to the power-balancing of smart phones. The results also indicates that although AHCPN may consume extra energy because of sending smart packets periodically, it can effectively reduce the number of drained smart phones by balancing the remaining battery power of smart handsets.

\section{Conclusions}
\label{Conclu}

In this paper we propose an infrastructure-less indoor emergency response system to evacuate civilians with the aid of smart handsets and cloud servers. A coordinated emergency navigation algorithm is proposed to guide evacuees in loose groups. The experimental results prove that the algorithm can increase the survival rate by reducing the number of drained smart phones in an evacuation process and raising the likelihood for an evacuee with a depleted mobile device to encounter other evacuees in the line of sight and follow them to egresses. Due to the considerable energy consumption between the Cloud and the smart phones during communication processes, an AHCPN based energy efficient protocol is also presented to prolong the life time of smart handsets. Simulations indicate that the protocol can significantly decrease the number of drained smart phones in an evacuation process and balance the remaining battery power among portable devices.

\bibliographystyle{spphys}%
\bibliography{cloudcomputing}
\end{document}